\documentclass[12pt]{article}
\usepackage{amssymb}
\usepackage{epsfig}
\usepackage{float}
\usepackage{graphicx}
\usepackage{amsmath}
\newcommand{\nua}[1]{\ensuremath{\rlap
           {\kern-2.5pt\ensuremath
           {\overset{\scriptscriptstyle(-)}{\phantom{\nu}}}}
           {\ensuremath{{\nu}_{#1}}}}}

\begin{document}
\begin{center}
{\bf  Neutrino Masses from the Point of View of Economy and Simplicity}
\end{center}

\begin{center}
S.  Bilenky
\end{center}
\begin{center}
{\em  Joint Institute for Nuclear Research, Dubna, R-141980,
Russia\\}
{\em TRIUMF
4004 Wesbrook Mall,
Vancouver BC, V6T 2A3
Canada\\}
\end{center}

\begin{abstract}
In the framework of such basic principles as local gauge invariance, unification of the weak and electromagnetic interactions and spontaneous symmetry breaking in the Standard Model   the most economical and simplest possibilities are realized. We discuss  the problem of neutrino masses  from the point of view of economy and simplicity. It is  unlikely that neutrino masses are of the same SM origin as masses of leptons and quarks. The Weinberg effective Lagrangian is the simplest and the most economical, beyond the Standard Model mechanism of the generation of  small Majorana neutrino masses. The resolution of the sterile neutrino anomaly and observation of the neutrinoless double $\beta$-decay would be crucial tests of this mechanism.

\end{abstract}

\section{Neutrino and the Standard Model}

\subsection{Introduction}
In 1957, soon after the discovery of the parity violation in the $\beta$-decay \cite{Wu:1957my} and $\mu$-decay \cite{Garwin:1957hc,Friedman:1957mz}, Landau \cite{Landau:1957tp}, Lee and Yang \cite{Lee:1957qr} and Salam \cite{Salam:1957st} proposed {\em the theory of the two-component neutrino}.

This theory was based on the assumption that neutrino mass is equal to zero. For massless neutrino  two-component left-handed (right-handed) neutrino fields $\nu_{L,R}(x)=\frac{1}{2}(1\mp\gamma_{5})~\nu(x)$ satisfy decoupled Weil equations
\begin{equation}\label{Weil}
i\gamma^{\alpha}\partial_{\alpha}~\nu_{L,R}(x)=0.
\end{equation}
Thus, for neutrino with equal to zero mass neutrino field can be $\nu_{L}(x)$ or $\nu_{R}(x)$.

The Weil equations, however, are not invariant under the space inversion.\footnote{The equations (\ref{Weil}) were proposed by H. Weil in 1929. However, during many years they were disregarded because they do not conserve parity. In the Pauli book "Quantum Mechanics`` \cite{Pauli} it was written "...because the equation for
$\psi_{L}(x)$ ($\psi_{R}(x))$ is not invariant under space reflection it is not applicable to the physical reality".} When large violation of parity was discovered in the weak interaction \cite{Wu:1957my,Garwin:1957hc,Friedman:1957mz} the fields $\nu_{L}(x)$ and $\nu_{R}(x)$ became natural candidates for the neutrino field (two-component neutrino theory \cite{Landau:1957tp,Lee:1957qr,Salam:1957st}).

The field $\nu_{L}(x)$ ($\nu_{R}(x)$) is the field of neutrino with negative (positive) helicity and antineutrino with positive (negative) helicity. Neutrino helicity was measured in the classical Goldhaber at al. experiment \cite{Goldhaber:1958nb} Goldhaber et al. concluded: ``our result is compatible with  100\% negative helicity  of neutrino." Thus from the experiment \cite{Goldhaber:1958nb} it followed that neutrino was left-handed particle and the neutrino field was $\nu_{L}(x)$.

Let us stress that {\em the two-component field $\nu_{L}(x)$ is the  simplest, most economical possibility for massless neutrino: only two degrees of freedom} (dof), not four dof as in the general four-component Dirac case.

The theory of the two-component neutrino played a crucial role in the creation of the Feynman and Gell-Mann \cite{Feynman:1958ty},  Marshak and Sudarshan \cite{Sudarshan:1958vf} current $\times$ current, universal $V-
A$ theory of the weak interaction.

This successful phenomenological theory of the weak interaction was based on the assumption that all fields entered into the Lagrangian of the weak interaction in the same way as neutrino field: as left-handed components. Assuming  the universality of the weak interaction\footnote{Idea of the universality of the weak interaction was first discussed in \cite{Pontecorvo:1947vp}. Later the universality of the weak interaction was suggested in \cite{Puppi:1948qy,Yang:1950qm}.}  it was assumed that the Hamiltonian has the following current$\times$current form
\begin{equation}\label{VA}
 \mathcal{H}_{I}= \frac{G_{F}}{\sqrt{2}} ~ j^{\alpha} ~ j^{+}_{\alpha}.
\end{equation}
Here $G_{F}$ is the Fermi constant and
\begin{equation}\label{VA1}
 j^{\alpha}=2 ~(\bar{p}_L \gamma^{\alpha} n_L  +  \bar\nu_{eL}
\gamma ^{\alpha} e_L  +  \bar\nu_{\mu L}  \gamma^{\alpha} \mu_L)
\end{equation}
is the charged weak current\footnote{Feynman and Gell-Mann,  Marshak and Sudarshan considered one type of neutrino. Later it was proved by the Brookhaven neutrino experiment \cite{Danby:1962nd}, which was proposed by B. Pontecorvo \cite{Pontecorvo:1959sn}, that $\nu_{\mu}$ and $\nu_{e}$ are different particles.}

Let us stress that (\ref{VA}) is {\em the simplest possible phenomenological Hamiltonian of the weak interaction in the low energy region.}\footnote{The general phenomenological Hamiltonian of the $\beta$-decay is characterized by 10 coupling constants. If we include $\mu$-processes, assume that left-handed components of all fields enter into the Hamiltonian and postulate universality of the weak interaction we come to the Hamiltonian which is characterized by only one (Fermi) constant. Feynman and Gell-Mann \cite{Feynman:1958ty}   ``... have adopted the point of view  that the weak interactions arise from the interaction of a current $j_{\alpha}$ with itself, possibly via an intermediate charged vector meson  of  high mass."}

\subsection{Standard Model Interaction (leptons and neutrinos)}
The two-component neutrino theory and V-A theory were  crucial for the creation of
the $SU_{L}(2)\times U_{Y}(1)$ unified theory of the weak and electromagnetic interactions, the Standard Model \cite{Glashow:1961tr,Weinberg:1967tq,Salam:1968rm}.

The Standard Model is based on the following general principles
\begin{enumerate}
  \item Local gauge $SU_{L}(2)\times U_{Y}(1)$ invariance.

   \item Unification of the weak and electromagnetic interactions.

  \item Brout-Englert-Higgs mechanism of mass generation \cite{Higgs:1964pj,Higgs:1964ia,Higgs:1966ev,Englert:1964et}.

\end{enumerate}
In the framework of these general principles {\em  the simplest, most economical  possibilities are realized in the Standard Model.}\footnote{Interesting that when the Standard Model was proposed it was considered more as a correct strategy than a realistic model of the electroweak interaction (see \cite{Weinberg:1979pi}). It occurred that the first simplest model was the correct one.}

In fact, $SU_{L}(2)$ is \emph{the minimal group} which allow to unify two-component, left-handed,  neutrino fields $\nu'_{lL}(x)$
and left-handed,  lepton fields $l'_{L}(x)$  into
lepton doublets\footnote{The mining of primes  will be clear later.}
\begin{equation}\label{SU}
    \psi_{lL}(x)=\left(
\begin{array}{c}
\nu'_{lL}(x) \\
l'_L(x) \\
\end{array}
\right),\quad l=e,\mu,\tau.
\end{equation}
{\em The minimal group} which allows to unify the weak and electromagnetic interactions is $SU_{L}(2)\times U_{Y}(1)$ local group, where $U_{Y}(1)$ is the group of the hypercharge $Y$ determined by the Gell-Mann-Nishijima relation
\begin{equation}\label{GM-N}
Q=I_{3}+\frac{1}{2}Y,
\end{equation}
where $Q$ is the electric charge (in the units of the proton charge) and $I_{3}$
is the third projection of the $SU(2)_{L}$ isotopic spin.

The SM interaction Lagrangian
\begin{equation}\label{SM1}
\mathcal{L}_{I}(x)=-g~\vec{j}_{\alpha}(x)\cdot\vec{A}^{\alpha}(x)
-g'\frac{1}{2}j^{Y}_{\alpha}(x)B^{\alpha}(x).
\end{equation}
is {\em the minimal interaction Lagrangian} which satisfies the requirements of the local $SU_{L}(2)\times U_{Y}(1)$ invariance.

Here $g$ and $g'$ are $SU_{L}(2)$ and $U_{Y}(1)$ dimensionless coupling constants, $\vec{A}^{\alpha}(x)$ and $B^{\alpha}(x)$ are, correspondingly, $SU_{L}(2)$ and $U(1)_{Y}$  vector gauge fields and the hypercurrent $j^{Y}_{\alpha}(x)$ is given by the expression
\begin{equation}\label{SM2}
\frac{1}{2}j^{Y}_{\alpha}(x)=-j^{3}_{\alpha}(x)+j^{EM}_{\alpha}(x),
\end{equation}
where
\begin{equation}\label{SM3}
j^{EM}_{\alpha}(x)=(-1)(\sum_{l}\bar l'_{L}(x)\gamma_{\alpha}l'_{L}(x)+
\sum_{l}\bar l'_{R}(x)\gamma_{\alpha}l'_{R}(x))=(-1)\sum_{l}\bar l'(x)\gamma_{\alpha}l'(x)
\end{equation}
is the electromagnetic lepton current.

The interaction Lagrangian $\mathcal{L}_{I}(x)$ is  a sum of the charged current (CC), neutral current (NC) and electromagnetic (EM) interactions
\begin{equation}\label{SM4}
\mathcal{L}_{I}=(-\frac{g}{2\sqrt{2}}j^{CC}_{\alpha}W^{\alpha}+
\mathrm{h.c.}) -\frac{g}{2\cos\theta_{W}}j^{NC}_{\alpha}Z^{\alpha} -ej^{EM}_{\alpha}A^{\alpha}.
\end{equation}
Here
\begin{equation}\label{SM5}
j^{CC}_{\alpha}=2(j^{1}_{\alpha} +ij^{2}_{\alpha})=2\sum_{l=e,\mu,\tau}\bar\nu'_{lL}\gamma_{\alpha}l'_{L}
\end{equation}
is the lepton charged current,
\begin{equation}\label{SM6}
j^{NC}_{\alpha}=2j^{3}_{\alpha}-2\sin^{2}\theta_{W}j^{EM}_{\alpha}
\end{equation}
is the neutral current, $j^{EM}_{\alpha}$ is the electromagnetic current,
\begin{equation}\label{SM7}
W_{\alpha}=\frac{1}{\sqrt{2}}(A^{1}_{\alpha}-i A^{2}_{\alpha})
\end{equation}
is the field of the vector charged $W^{\pm}$ bosons and
\begin{equation}\label{SM8}
Z_{\alpha}=\cos\theta_{W}A^{3}_{\alpha}-\sin\theta_{W}B_{\alpha},~~~
A_{\alpha}=\sin\theta_{W}A^{3}_{\alpha}+\cos\theta_{W}B_{\alpha}
\end{equation}
are the field of the vector neutral $Z^{0}$ bosons and the electromagnetic field, respectively.

The weak (Weinberg) angle $\theta_{W}$ is determined by the relation
\begin{equation}\label{SM9}
    \tan\theta_{W}=\frac{g'}{g}
\end{equation}
and the electric charge $e$ is given by
\begin{equation}\label{SM10}
 e=g\sin\theta_{W}.
\end{equation}
The Standard Model is based on the  Brout-Englert-Higgs mechanism of  spontaneous symmetry breaking \cite{Higgs:1964pj,Higgs:1964ia,Higgs:1966ev,Englert:1964et}. Before the spontaneous symmetry breaking there are no mass terms in the Lagrangian.

In order to generate masses of $W^{\pm}$ and  $Z^{0}$ vector bosons we need three Goldstone degrees of freedom. One Higgs $SU_{L}(2)$ doublet
\begin{eqnarray}
\phi(x)=\left(
\begin{array}{c}
\phi_{+}(x) \\
\phi_{0}(x)\\
\end{array}
\right),
\label{Higgs}
\end{eqnarray}
where $\phi_{+}(x)$ and $\phi_{0}(x)$ are   complex charged and neutral scalar fields (4 dof), is {\em the most economical possibility}.
We can choose (unitary gauge)
\begin{eqnarray}\label{1Higgs}
\phi(x)=\left(
\begin{array}{c}
0 \\
\frac{v+H(x)}{\sqrt{2}}\\
\end{array}
\right)
\end{eqnarray}
where $v=\mathrm{const}$ is a vacuum expectation value (vev) of the Higgs field and $H(x)=H^{\dag}(x)$ is a field of scalar, neutral Higgs particles. With the choice (\ref{1Higgs}) the symmetry will be spontaneously broken.

Thus, the Standard Model {\em predicted existence of a neutral scalar Higgs boson.} Discovery of the Higgs boson at LHC \cite{Chatrchyan:2012xdj,Aad:2012tfa} perfectly confirmed this prediction.

Masses of $W^{\pm}$ and $Z^{0}$ bosons are given by the relations
\begin{equation}\label{Higgs22}
m_{W}=\frac{1}{2}g~v,~~~
m_{Z}=\frac{1}{2}\sqrt{g^{2}+g^{'2}}~v=\frac{g}{2\cos\theta_{W}}~v.
\end{equation}
For the Fermi constant $G_{F}$ we have
\begin{equation}\label{Fermi}
    \frac{G_{F}}{\sqrt{2}}=\frac{g^{2}}{8m^{2}_{W}}.
\end{equation}
From (\ref{Higgs22}) and (\ref{Fermi}) it follows that
\begin{equation}\label{vev1}
    v=(\sqrt{2}G_{F})^{-1/2}\simeq 246~\mathrm{GeV}.
\end{equation}

\subsection{Dirac Mass Term of Charged Leptons}

Charged leptons (and quarks) are Dirac particles (particles and antiparticles differ by charges). The Dirac mass term is a Lorentz-invariant product of left-handed and right-handed fields. It is generated by the Yukawa interaction.

 The Lagrangian of the $SU_{L}(2)\times U_{Y}(1)$ invariant Yukawa interaction of the charged lepton and Higgs fields is given by the expression
\begin{equation}\label{Yukawa}
    \mathcal{L}^{Y}_{I}(x)=-\sqrt{2}\sum_{l_{1},l_{2}}\bar\psi_{l_{1}L}(x)
    Y_{l_{1}l_{2}}l'_{2R}(x)\phi(x)+\mathrm{h.c.}
\end{equation}
Here
\begin{eqnarray}\label{Yukawa2}
\psi_{lL}(x)=\left(
\begin{array}{c}
\nu'_{lL}(x) \\
l'_{L}(x)\\
\end{array}
\right),\quad l=e,\mu,\tau
\end{eqnarray}
is the lepton doublet, $l'_{R}$ is the  singlet right-handed lepton field and $Y$ is a dimensionless complex $3\times3$ matrix which is not constrained by the $SU_{L}(2)\times U_{Y}(1)$ symmetry.

After the spontaneous symmetry breaking from (\ref{1Higgs}) and (\ref{Yukawa})  we find
\begin{equation}\label{Yukawa3}
\mathcal{L}^{Y}_{I}(x)=-\sum_{l=e,\mu,\tau}m_{l}~\bar l(x)~l(x)
-\sum_{l=e,\mu,\tau}f_{l}~\bar l(x)~l(x).
\end{equation}
Here
\begin{equation}\label{Yukawa4}
    m_{l}=y_{l}~v
\end{equation}
is the mass of the lepton $l$ ($l=e,\mu,\tau$), $y_{l}$ is  the Yukawa coupling (an eigenvalue of the matrix $Y$)   and $l(x)=l_{L}(x)+l_{R}(x)$ is the field of the lepton with the mass $m_{l}$.

Notice that the second term of the Lagrangian (\ref{Yukawa3}) is the Lagrangian of interaction of charged leptons and the Higgs boson. The Standard Model predicts the constants of this interaction $f_{l}$:
\begin{equation}\label{Yukawa5}
f_{l}\equiv y_{l} =\frac{m_{l}}{v}.
\end{equation}
The SM  also predicts constants of interaction of quarks and the Higgs boson
\begin{equation}\label{Yukawa5}
f_{q} =\frac{m_{q}}{v},\quad q=u,d,c,...
\end{equation}
These predictions are in a good agreement with recent LHC data (see \cite{Tanabashi:2018oca}).

The primed fields $l'_{L}(x)$ and  $l'_{R}(x)$  are connected with $l_{L}(x)$ and $l_{R}(x)$ by the relations
\begin{equation}\label{8qYuk}
l'_{L}(x)=\sum_{l=e,\mu,\tau}(V_{L})_{l'l}l_{L}(x),\quad
l'_{R}(x)=\sum_{l=e,\mu,\tau}(V_{R})_{l'l}l_{R}(x),
\end{equation}
where $V_{L}$ and $V_{R}$ are unitary matrices.

Notice that for charged particles both left-handed and right-handed fields enter into $SU_{L}(2)\times U_{Y}(1)$ invariant Lagrangian of the Standard Model (right-handed fields enter into the electromagnetic and neutral currents). Thus, {\em the Yukawa interactions} which generate  the Dirac mass terms  {\em do not require additional degrees of freedom}.

\subsection{Neutrino Masses in the Standard Model}

Let us assume that  the following $SU(2)_{L}\times U(1)_{Y}$ invariant Yukawa interaction
\begin{equation}\label{YukawaNu}
    \mathcal{L}^{Y\nu}_{I}(x)=-\sqrt{2}\sum_{l_{1},l_{2}}\bar\psi_{l_{1}L}(x)
    Y^{\nu}_{l_{1}l_{2}}\nu'_{l_{2}R}(x)\tilde{\phi}(x)+\mathrm{h.c.}
\end{equation}
enters into the total Lagrangian. Here $\nu'_{l R}$ are $SU(2)_{L}$ singlets, $\tilde{\phi}=i\tau_{2}~\phi^{*}$ is the conjugated Higgs doublet and $Y^{\nu}$ is a $3\times3$ dimensionless matrix.

After the spontaneous symmetry breaking from (\ref{1Higgs}) and (\ref{YukawaNu}) we come to the neutrino mass term
\begin{equation}\label{YukawaNu1}
  \mathcal{L}^{m}(x)=-\sum^{3}_{i=1}m_{i}~\bar\nu_{i}(x) \nu_{i}(x).  \end{equation}
Here
\begin{equation}\label{YukawaNu2}
    m_{i}=y^{\nu}_{i}~v,
\end{equation}
where $y^{\nu}_{i}$ is an eigenvalue of the matrix $Y^{\nu}$. From (\ref{YukawaNu1}) follows that $\nu_{i}(x)$ is the  field of neutrino with mass $m_{i}$.

The primed neutrino fields $\nu'_{lL}(x)$ are connected with the fields $\nu_{iL}(x)$ by the relation
\begin{equation}\label{YukawaNu3}
\nu'_{l}(x)=\sum_{i=1,2,3}(U_{L})_{li}~\nu_{iL}(x),\quad l=e,\mu,\tau
\end{equation}
where $U_{L}$ is an unitary matrix.

It is obvious that the total Lagrangian, in which  neutrino mass term (\ref{YukawaNu1}) enter, is invariant under the global transformation
\begin{equation}\label{Glob}
\nu_{i}(x)\to e^{i\Lambda}\nu_{i}(x),\quad l(x)\to e^{i\Lambda}l(x),\quad q(x)\to  q(x),
\end{equation}
where $\Lambda$ is an arbitrary constant. From  invariance under the transformation (\ref{Glob}) follows that the total lepton number $L$  is conserved and {\em $\nu_{i}(x)$ is the field of Dirac neutrinos} (neutrino $\nu_{i}$ and antineutrino $\bar\nu_{i}$ are different particles: $L(\nu_{i})=1, L(\bar\nu_{i})=-1$).

In spite, as we have seen, formally it is possible to generate Dirac neutrino masses in the same way as lepton (and quark) masses are generated, it is a common opinion that in the Standard Model neutrinos are massless two-component particles.

\begin{enumerate}
  \item  All fermion masses generated by the standard Higgs mechanism are given by products of corresponding Yukawa couplings and the  Higgs vev $v\simeq 246$ GeV (see (\ref{Yukawa4}) and (\ref{YukawaNu2})). Neutrino masses are much smaller than masses of leptons and quarks. Thus, if neutrino masses are of the SM origin, corresponding Yukawa couplings must be many orders of magnitude smaller than Yukawa couplings  of leptons and quarks.

 In fact, let us consider the third family. Yukawa couplings  of t and  b quarks and $\tau$-lepton are equal, respectfully,
\begin{equation}\label{YukawaNu10}
y_{t}\simeq 7\cdot 10^{-1},~~ y_{b}\simeq 2\cdot 10^{-2},~~y_{\tau}\simeq 7\cdot 10^{-3}.
\end{equation}
At present absolute values of neutrino masses are unknown. However, from existing neutrino  data we can obtain the following  estimate for the largest neutrino mass: $(5\cdot 10^{-2}\lesssim m_{3}\lesssim 3\cdot 10^{-1})$ eV. Thus, for the neutrino Yukawa coupling $y^{\nu}_{3}$ we have
\begin{equation}\label{YukawaNu12}
 2\cdot 10^{-13}\lesssim    y^{\nu}_{3}\lesssim \cdot 10^{-12}.
\end{equation}
From (\ref{YukawaNu10}) follow that  Yukawa couplings of quarks and lepton of the third generation  are in the range $\sim (1-10^{-2})$.
The neutrino Yukawa coupling is about
 ten orders of magnitude smaller. {\em We conclude that it is very unlikely that neutrino masses are of the same Standard Model origin as masses of leptons and quarks.}

\item
 The second argument in favor of massless neutrinos in the Standard Model is economy. In fact, into the SM Lagrangian which does not include  Yukawa interactions, enter left-handed and right-handed components  of  {\em charged} fields. Thus, Yukawa interactions, which generate mass terms of charged leptons and quarks, do not require additional degrees of freedom.

 Neutrinos have no direct electromagnetic interaction in which left-handed and  right-handed fields enter. In the interaction Lagrangian there are only left-handed neutrino fields. The Yukawa interaction (\ref{YukawaNu}), which generate mass term of neutrinos, requires to double the number of the neutrino degrees of freedom. Thus, {\em the most economical possibility is the Standard Model with massless neutrinos} (without  right-handed neutrino fields).

\end{enumerate}

\section{Beyond the SM Neutrino Masses}
\subsection{Introduction. Majorana Mass Term}
The SM charged current is given by the relation (\ref{SM5}). Taking into account
(\ref{8qYuk}) we have
\begin{equation}\label{5SM}
j^{CC}_{\alpha}(x)=2\sum_{l=e,\mu,\tau}\bar\nu_{lL}(x)\gamma_{\alpha}l_{L}(x).
\end{equation}
Here $l(x)$ is the field of the lepton $l$  with the mass $m_{l}$
($l=e,\mu,\tau$) and
\begin{equation}\label{flavf}
\nu_{lL}(x)=\sum_{l_{1}}(V^{\dag}_{L})_{ll_{1}}~\nu'_{l_{1}L}
\end{equation}
is {\em the flavor neutrino field}. As it is well known, the Standard Model with the current (\ref{5SM}) is in a perfect agreement with experiment.

The first neutrino mass term was introduced by Gribov and Pontecorvo \cite{Gribov:1968kq} many years ago. They showed that it
is  possible to build the neutrino mass term in which only left-handed  flavor fields $\nu_{lL}(x)$ entered. If fact, if we take into account that the conjugated field
\begin{equation}\label{conj}
(\nu_{lL})^{c}\equiv \nu_{lL}^{c}=C\bar\nu_{lL}^{T},
\end{equation}
($C$ is the matrix of the charge conjugation which satisfies the conditions ($C\gamma_{\alpha}^{T}C^{-1}=-\gamma_{\alpha},~C^{T}=-C$) is the right-handed field and assume that {\em the total lepton number $L$ is not conserved} for the neutrino mass term we have\footnote{ In \cite{Gribov:1968kq} the simplest case of two flavor neutrinos was considered and $\mathcal{L}^{\mathrm{M}}$ was interpreted as an additional (superweak) interaction between neutrinos. The Majorana mass term for a general case was discussed in \cite{Bilenky:1987ty}.}
\begin{equation}\label{Mj}
\mathcal{L}^{\mathrm{M}}=-\frac{1}{2}\sum_{l',l}\bar\nu_{l'L}M^{\mathrm{M}}_{l'l}\nu_{lL}^{c} +\mathrm{h.c.}
\end{equation}
Here $M^{\mathrm{M}}=(M^{\mathrm{M}})^{T}$ is a symmetrical $3\times3$ mixing matrix. The matrix $M^{\mathrm{M}}$ can be presented in the form
\begin{equation}\label{Mj1}
    M^{\mathrm{M}}=U~m~U^{T},
\end{equation}
where is an unitary matrix ($U^{\dag}U=1$)  and $m$ is a diagonal matrix ($m_{ik}=m_{i}\delta_{ik},~~m_{i}>0$).

From (\ref{Mj}) and (\ref{Mj1}) follows that
\begin{equation}\label{Mj2}
 \mathcal{L}^{\mathrm{M}}=-\frac{1}{2}\sum^{3}_{i=1}m_{i}~\bar\nu_{i}\nu_{i}.
\end{equation}
Here $\nu_{i}$ is the field of {\em the Majorana neutrino} with the mass $m_{i}$:
\begin{equation}\label{Mj3}
\nu_{i}=\nu^{c}_{i} =C\bar\nu^{T}_{i}.
\end{equation}
The flavor field $\nu_{lL}$ is the "mixed" field
\begin{equation}\label{Mj4}
    \nu_{lL} =\sum^{3}_{i=1}U_{li}\nu_{iL},
\end{equation}
where $U$ is PMNS \cite{Pontecorvo:1957qd,Maki:1962mu} mixing matrix. The mass term (\ref{Mj}) is called the Majorana mass term.

In conclusion we like to stress the following
\begin{itemize}
  \item The Majorana mass term (\ref{Mj}) is {\em the most economical mass term.} The flavor neutrino fields $\nu_{lL}$ enter into the interaction Lagrangian and into the neutrino mass term (the number of neutrino dof is minimal).
  \item In the pure phenomenological approach, we considered here, neutrino masses and mixing angles are parameters. There are no any clues why neutrino masses $m_{i}$ are much smaller then lepton and quarks masses.
\end{itemize}

\subsection{Weinberg Mechanism of the Neutrino Mass Generation}
Much more powerful approach to the problem of the neutrino masses is based on the method of the effective Lagrangian \cite{Weinberg:1979sa}, a general method which allows to describe effects of a beyond the Standard Model physics. The effective Lagrangian is a dimension five or more non renormalizable  Lagrangian, invariant under   $SU(2)_{L}\times U(1)_{Y}$  transformations and  built from the Standard Model fields.

In order to built the effective Lagrangian which generate a
neutrino mass term let us consider  the $SU_{L}(2)\times U_{Y}(1)$  invariant
\begin{equation}\label{effL}
  (\tilde{\phi }^{\dag}~ \psi_{lL})~~~(l=e,\mu,\tau).
\end{equation}
which has dimension $M^{5/2}$. In (\ref{effL})
$\psi_{lL}$ is the lepton doublet given by (\ref{Yukawa2}) and $\tilde{\phi }$ is the conjugated Higgs doublet.

After spontaneous symmetry breaking we have
\begin{equation}\label{effL1}
(\tilde{\phi }^{\dag}~ \psi_{lL})\to \frac{v}{\sqrt{2}}~\nu'_{lL}.
\end{equation}
It is obvious from (\ref{effL1}) that {\em the only effective Lagrangian}  which generates the neutrino mass term is given by the expression
\begin{equation}\label{effL2}
\mathcal{L}_{I}^{\mathrm{eff}}=-\frac{1}{\Lambda}~\sum_{l_{1},l_{2}}(\bar \psi_{l_{1}L}\tilde{\phi })~ X'_{l_{1}l_{2}}~(\tilde{\phi }^{T} \psi_{l_{2}L}^{c})+\mathrm{h.c.}
\end{equation}
first considered in  \cite{Weinberg:1979sa}. Here $X'$ is a $3\times 3$ dimensionless, symmetrical matrix. The operator in (\ref{effL2}) has dimension $M^{5}$.

We introduced in (\ref{effL2}) the coefficient $\frac{1}{\Lambda}$ where $\Lambda$ has dimension $M$. The constant $\Lambda$  characterizes a scale of a beyond the SM physics.

{\em The effective Lagrangian (\ref{effL2}) does not conserve the total lepton number $L$.} Let us notice that the global invariance and conservation of $L$ (and $B$) is not a fundamental symmetry of a Quantum Field Theory. This is connected with the fact that constant phase is not a dynamical variable. In the Standard Model local gauge symmetry and renormalizability  ensure conservation of $L$. It is natural to expect that a beyond the Standard Model theory does not conserve $L$  (see \cite{Weinberg:1980bf,Witten:2017hdv}).

After the spontaneous symmetry breaking from (\ref{effL2}) we come to the following Majorana mass term
\begin{equation}\label{2effL}
\mathcal{L}^{\mathrm{M}}= -\frac{1}{2}\,\frac{v^{2}}{\Lambda}\sum_{l_{1},l_{2}}
\bar\nu'_{l_{1}L}\,X_{l_{1}l_{2}} \nu'^{c}_{l_{2}L}+\mathrm{h.c.}
\end{equation}
Taking into account (\ref{flavf}), we can rewrite the mass term (\ref{2effL}) in the usual form
\begin{equation}\label{2effLa}
\mathcal{L}^{\mathrm{M}}= -\frac{1}{2}\,\sum_{l_{1},l_{2}}
\bar\nu_{l_{1}L}\,M^{\mathrm{M}}_{l_{1}l_{2}}  \nu^{c}_{l_{2}L}+\mathrm{h.c.}
\end{equation}
where the Majorana mass matrix is given by the relation
\begin{equation}\label{Majmat}
M^{\mathrm{M}}=\frac{v^{2}}{\Lambda}~ X
\end{equation}
and $X=V^{\dag}_{L}~X'~(V^{\dag}_{L})^{T}$.

We have
\begin{equation}\label{effL3}
X=U~x~U^{T},
\end{equation}
where $U~U^{\dag}=1$  and $x_{ik}=x_{i}\delta_{ik},~~x_{i}>0$.

From (\ref{2effLa}), (\ref{Majmat}) and (\ref{effL3}) it follows that the Majorana mass term (\ref{2effLa}) takes the following standard form
\begin{equation}\label{effL4}
\mathcal{L}^{\mathrm{M}}=-\frac{1}{2}\sum^{3}_{i=1}m_{i}~\bar \nu_{i}\nu_{i}.
\end{equation}
Here
\begin{equation}\label{effL5}
\nu_{i}= \nu^{c}_{i}
\end{equation}
is the field of the Majorana neutrino with the mass
\begin{equation}\label{effL6}
m_{i}=\frac{v^{2}}{\Lambda}~x_{i}.
\end{equation}
The flavor neutrino field $\nu_{lL}$ is connected with fields of the Majorana neutrinos with definite masses by the standard mixing relation
\begin{equation}\label{effL7}
\nu_{lL}=\sum^{3}_{i=1}U_{li}~\nu_{iL}
\end{equation}
 {\em The effective Lagrangian approach to neutrino masses allows to understand a possible origin of the  smallness of neutrino masses}. In fact, if the factor
\begin{equation}\label{effL12}
\frac{v}{\Lambda}=\frac{\mathrm{EW~scale}}{\mathrm{scale~of~a~new~ physics}}\ll 1
\end{equation}
in this case neutrino masses, given by the relation  (\ref{effL6}), are much smaller than proportional to $v$ Standard Model masses of charged leptons and quarks.

 Let us stress, however, that from the values of neutrino masses, without knowledge of the dimensionless parameters $x_{i}$, we can not estimate the scale of a new  $L$-violating physics.

In fact, from existing neutrino data we can estimate that the largest neutrino mass $m_{3}$ is in the range \footnote{The lower bound can be obtained from neutrino oscillation data if we assume normal ordering of the neutrino masses ( $m_{3}>\sqrt{\Delta m^{2}_{A}}\simeq  5\cdot 10^{-2}~\mathrm{eV}$, where $\Delta m^{2}_{A}$ is the atmospheric neutrino mass-squared difference). The upper bound in (\ref{m3}) can be obtained from a conservative cosmological bound
$\sum m_{i}\lesssim 1$ eV.}
\begin{equation}\label{m3}
 5\cdot 10^{-2}\lesssim m_{3}\lesssim 3\cdot 10^{-1}~\mathrm{eV}.
\end{equation}
From (\ref{effL6}) and (\ref{m3}) we have the relation
\begin{equation}\label{effL13}
\Lambda \simeq (2\cdot 10^{14}-10^{15})~x_{3}~\mathrm{GeV}.
\end{equation}
From this relation we can conclude that if $x_{3}\simeq 1$ (like square of the Yukawa coupling of the t-quark) in this case $\Lambda$ is of the order of a GUT scale
\begin{equation}\label{effL14}
\Lambda \simeq (2\cdot 10^{14}-10^{15})~\mathrm{GeV}.
\end{equation}
If we assume that $\Lambda \simeq 1~\mathrm{TeV}$ (a scale which can be reached at LHC) in this case $x_{3}$ has a very small (unphysical) value in the range
\begin{equation}\label{effL15}
x_{3}\simeq (1-5)\cdot 10^{-12}.
\end{equation}
Summarizing, there is unique, non renormalizable, $SU_{L}(2)\times U_{Y}(1)$  invariant, beyond the SM effective Lagrangian which generate the neutrino mass term. This Lagrangian does not conserve the total lepton number $L$ and has the minimal dimension five. From the requirements of the invariance it follows that Higgs field
quadratically enters into the Lagrangian. This leads to quadratic dependence of beyond the SM neutrino masses on the parameter $v$ (in contrast to linear dependence of SM masses on $v$). From dimensional arguments follows that in the expression for neutrino masses must enter the term $\frac{1}{\Lambda}$, where $\Lambda$ characterizes the scale of a beyond the SM physics which, apparently, is much larger than the electroweak scale $v$. {\em The factor $\frac{v}{\Lambda}$ naturally ensures smallness of neutrino masses.}

\subsection{On the origin of the Weinberg effective Lagrangian}
There exist many models in which the Weinberg effective Lagrangian is generated.
The simplest and most economical approach is based on the assumption that exist
heavy Majorana leptons $N_{i}$, $SU_{L}(2)$ singlets, which interact with
lepton-Higgs pairs via the lepton number violating  Yukawa Lagrangian \cite{Weinberg:1979sa}
\begin{equation}\label{Leff}
\mathcal{L}_{I}=-\sqrt{2}\sum_{l, i}(\bar \psi_{l L}\tilde{\phi })y_{li}~N_{iR}+\mathrm{h.c.}
\end{equation}
Here
\begin{equation}\label{1Leff}
N_{i}=N^{c}_{i}=C(\bar N_{i})^{T}, ~~~i=1,2,...
\end{equation}
is the field of the Majorana heavy lepton with the mass $M_{i}$,
$\psi_{l L}$ and $\tilde{\phi}$ are SM lepton and conjugated Higgs doublets and $y_{li}$ are Yukawa couplings.

For low-energy processes with virtual heavy leptons from (\ref{Leff}) in the tree approximation we obtain the Weinberg effective Lagrangian.\footnote{Classical example of the effective Lagrangian is
the famous dimension six  current$\times$current  Lagrangian of the weak interaction
\begin{equation}\label{nonumber}
\mathcal{L}_{I}=-\frac{G_{F}}{\sqrt{2}}j^{CC\alpha}~(j_{\alpha}^{CC})^{\dag}.
\end{equation}
which describe low energy CC weak processes  at $Q^{2}\ll m^{2}_{W}$. In the second order of the perturbation theory this Lagrangian is generated by the Standard Model  CC Lagrangian
\begin{equation}\label{nonumber}
 \mathcal{L}^{CC}_{I}=-\frac{g}{2\sqrt{2}}j^{CC}_{\alpha}W^{\alpha}+\mathrm{h.c.}   \end{equation}}
In fact, at $Q^{2}\ll M^{2}_{i}$ the propagator of the Majorana lepton  is given by the expression
\begin{equation}\label{2Leff}
 \langle 0|T(N_{iR}(x_{1})N^{T}_{iR}(x_{2}))|0\rangle=i\frac{1}{M_{i}}
\delta(x_{1}-x_{2})\frac{1+\gamma_{5}}{2}C.
\end{equation}
From (\ref{Leff}) and (\ref{2Leff})  we obtain the effective Lagrangian (\ref{effL2}) in which
\begin{equation}\label{3Leff}
\frac{1}{\Lambda}X'_{l_{1}l_{2}}=\sum_{i}y_{l_{1}i}\frac{1}{M_{i}}y_{l_{2}i}.
\end{equation}
Thus, a scale of a new physics $\Lambda$ is determined by masses of heavy Majorana leptons.

The mechanism, we considered,  is equivalent to the famous seesaw mechanism of the neutrino mass generation \cite{Minkowski:1977sc,GellMann:1980vs,Yanagida:1979as,Glashow:1979nm,
      Mohapatra:1980yp}. It is usually called the type I seesaw mechanism

There exist two additional possibilities to generate the effective Lagrangian  (\ref{effL2}) in the tree-approximation.
\begin{itemize}
  \item If exist {\em heavy triplet scalar bosons which interact with lepton pair and the Higgs pair}  the effective Lagrangian $\mathcal{L}_{\rm{eff}}$ will be generated by the exchange  of a virtual scalar boson between lepton and Higgs pairs (this scenario is called the type II seesaw mechanism).

 \item If exist {\em heavy Majorana triplet leptons which interact with lepton-Higgs  pair}, their exchange between  lepton-Higgs  pairs will generate the effective Lagrangian (\ref{effL2}) (the type III seesaw mechanism).

\end{itemize}

\subsection{ General Remarks}
The effective Lagrangian (seesaw) mechanism
is an attractive and economical beyond the Standard Model mechanism of the neutrino mass generation.

 \begin{enumerate}
  \item It allows to explain not only the smallness of neutrino masses but also it opens a possibility to explain the barion asymmetry of the Universe. In fact, if heavy Majorana leptons exist they   could be created in the early Universe. Their $CP$-violating decays  into Higgs-lepton pairs produce the lepton asymmetry of the Universe which via sphaleron processes generates the barion asymmetry (baryogenesis through leptogenesis, see reviews \cite{Buchmuller:2004tu,Davidson:2008bu,DiBari:2012fz}).

  \item The Weinberg effective Lagrangian has a  dimension five which is the {\em minimal dimension} for nonrenormalizable effective Lagrangians. For comparison an effective Lagrangian, responsible for $B$-violating proton decay, has dimension six. This means that investigation  of effects of small neutrino masses (via observation of neutrino oscillations and other effects) is, apparently, the best way to probe a beyond the Standard Model  physics at a very large scales.

\end{enumerate}
However, if the scale of a new physics $\Lambda $ is much larger than the electroweak scale $v$ the validity of this mechanism can not be directly tested.

There exist numerous   radiative neutrino mass models which lead  to the Weinberg effective Lagrangian. In these models values of neutrino masses $m_{i}$ are suppressed by  loop mechanisms which require existence of different  beyond the Standard Model particles with masses which could be much smaller than the GUT scale (for recent reviews see \cite{Cai:2017jrq,Everett2018}).

We could summarize the previous discussion in the following way. The major puzzle of neutrino masses, their extreme smallness, could be explained   by the Weinberg (seesaw) and many other beyond the SM mechanisms of neutrino mass generation. It is evident that without additional information on a beyond the SM physics we can not conclude what mechanism is realized in nature.

A wide class of minimal models (models without right-handed light neutrinos) lead to the Weinberg effective Lagrangian and the most economical Majorana mass term. In spite values of neutrino masses $m_{i}$ can not be predicted at present, from the structure of the Majorana mass term  we can come to the following general conclusions

\begin{enumerate}
  \item Neutrino with definite masses $\nu_{i}$ are Majorana particles.
  \item {\em The number of massive neutrinos is equal to the number of the lepton families (three)}.
  \item The neutrino mixing has the form
\begin{equation}\label{3nu}
 \nu_{lL}=\sum^{3}_{i=1}U_{li}\nu_{iL},\quad U^{\dag}U=1,
 \end{equation}
where $\nu_{i}$ is the field of the Majorana neutrino with the mass $m_{i}$ and $U$ is the unitary PMNS matrix.

 \end{enumerate}
As it is well known, the most sensitive experiments, which allow us to probe the Majorana nature of $\nu_{i}$, are experiments on the search for neutrinoless double $\beta$- decay ($0\nu\beta\beta$-decay) of some even-even nuclei. If this decay will be observed it would be important to check whether the decay is induced by three massive neutrinos. We will briefly discuss the status of the search for the $0\nu\beta\beta$-decay.

The crucial test of the idea of economy and simplicity can be realized in experiments on the search for forbidden by the Majorana mass term transitions of flavor neutrinos into sterile states. As it is well known,  some indications in favor of such transitions were obtained in LSND, MiniBooNE, reactor and Gallium short baseline neutrino experiments (see, for example \cite{Gariazzo:2018mwd}).
In many modern neutrino experiments these indications will be thoroughly checked.
The present status of the experiments on the search for sterile neutrino will be briefly discussed later.

\section{On Neutrino Oscillations}
\subsection{Neutrino States}
Discovery of neutrino oscillations  opened an era of investigation of neutrino masses and mixing. In this section we will briefly discuss this phenomenon (see, for example, \cite{Bilenky:2018hbz}).

In the case of the Majorana mass term the neutrino mixing has the form (\ref{3nu}).
Notice that neutrino mixing has the same form also in the case of the Dirac mass term. In this case the total lepton number is conserved and $\nu_{i}(x)$ are fields of the Dirac neutrinos. Let us stress that observable  neutrino transition probabilities do not depend on the nature of neutrinos with definite masses (Majorana or Dirac) \cite{Bilenky:1980cx,Langacker:1986jv}.

The state of flavor neutrino $\nu_{l}$ with momentum $\vec{p}$, which is produced in a weak decay together with lepton $l^{+}$, is given by a coherent superposition \cite{Bilenky:2001yh}
\begin{equation}\label{state}
  |\nu_{l}\rangle=\sum^{3}_{i=1}U^{*}_{li}~|\nu_{i}\rangle
\end{equation}
Here $|\nu_{i}\rangle$ is the state of neutrino with momentum
$\vec{p}$, mass $m_{i}$ and the energy $E_{i}\simeq p +\frac{m^{2}_{i}}{2E}$.

This relation is valid if the inequality
\begin{equation}\label{Heirel}
    \frac{2E}{|\Delta m_{ki}^{2}|}\gtrsim d
\end{equation}
is satisfied. Here $E$ is the neutrino energy, $d$ is a microscopic size of a neutrino source and $\Delta m_{ki}^{2}= m_{i}^{2}-m_{k}^{2}$.

The inequality (\ref{Heirel}) follows from the Heisenberg uncertainty relation. If (\ref{Heirel}) is satisfied, it is impossible to resolve emission of neutrinos with different masses in weak decays.\footnote{In current
neutrino oscillation experiments the condition (\ref{Heirel}) is perfectly satisfied. In fact, in the reactor KamLAND experiment ($E\simeq 4$ MeV, $\Delta m^{2}_{S}\simeq 8\cdot 10^{-5}~\mathrm{eV}^{2}$) we have $\frac{2E}{\Delta m_{S}^{2}}\simeq 2\cdot 10$ km.
In  long baseline accelerator experiments ($E \simeq 1$ GeV, $\Delta m^{2}_{A}\simeq 2.5\cdot 10^{-3}~\mathrm{eV}^{2}$) we find $\frac{2E}{\Delta m_{A}^{2}}\simeq 1.6\cdot 10^{2}$ km etc.}

We will  consider the general case of neutrino mixing and assume that in the neutrino mass term enter not only three current flavor fields $\nu_{lL}(x)$ ($l=e,\mu,\tau$) but also $n_{st}$ {\em sterile  fields} $\nu_{sL}(x)$ ($s=s_{1},...s_{n_{st}}$), the fields which do not enter in the Standard Model interaction Lagrangian.\footnote{This  terminology was introduced by B. Pontecorvo in 1967.} For the neutrino mixing we have in this case
\begin{equation}\label{Mix1}
\nu_{lL}(x)=\sum^{3+n_{st}}_{i=1}U_{li}~\nu_{iL}(x), \quad (l=e,\mu,\tau)
\end{equation}
and
\begin{equation}\label{Mix4}
  \nu_{sL}(x)=\sum^{3+n_{st}}_{i=1}U_{si}~\nu_{iL}(x)~~(s=s_{1},...s_{n_{st}}).
\end{equation}
Here $\nu_{i}(x)$ is the field of neutrino  with mass $m_{i}$ and $U$ is an $(3+n_{st})\times (3+n_{st})$ unitary mixing matrix.

If for all $3+n_{st}$ neutrino masses the inequality (\ref{Heirel}) is satisfied, for the state of the flavor neutrinos we have
\begin{equation}\label{States}
  |\nu_{l}\rangle=\sum^{3+n_{st}}_{i=1}U^{*}_{li}~|\nu_{i}\rangle.
\end{equation}
Analogously,  sterile states are determined by the following relations
\begin{equation}\label{States1}
  |\nu_{s}\rangle=\sum^{3+n_{st}}_{i=1}U^{*}_{si}~|\nu_{i}\rangle.
\end{equation}
From the unitarity of the mixing matrix we have
\begin{equation}\label{}
\langle \nu_{l'}|\nu_{l}\rangle=\delta_{l'l},~~\langle \nu_{l}|\nu_{s}\rangle=0,~~
\langle \nu_{s'}|\nu_{s}\rangle=\delta_{s's}.
\end{equation}

\subsection{Neutrino Transitions in Vacuum}
If at $t=0$ in a weak process the flavor neutrino $\nu_{l}$ with momentum $\vec{p}$ was produced, at the time $t$ the vacuum neutrino state is  given by
\begin{equation}\label{Prob1}
|\nu_{l}\rangle_{t}=e^{-iH_{0}t}|\nu_{l}\rangle,
\end{equation}
where $H_{0}$ is the free Hamiltonian. In neutrino experiments via observation of CC processes flavor neutrinos $\nu_{l'}$ ($l'=e,\mu,\tau$) are detected. For the probability of the $\nu_{l}\to \nu_{l'}$ transition we have
\begin{equation}\label{Prob2}
 \mathrm{P}(\nu_{l}\to \nu_{l'})=\left|\langle\nu_{l'}|e^{-iH_{0}t}|\nu_{l}\rangle\right|^{2}.
\end{equation}
Taking into account that $H_{0}|~\nu_{i}\rangle=E_{i}|~\nu_{i}\rangle$ we have
\begin{equation}\label{Prob3}
\mathrm{P}(\nu_{l}\to \nu_{l'})=|\sum^{3+n_{st}}_{i=1}\langle\nu_{l'}|\nu_{i}\rangle e^{-iE_{i}t}\langle\nu_{i}|\nu_{l}\rangle|^{2}=|\sum^{3+n_{st}}_{i=1}
U_{l'i}e^{-iE_{i}t}U^{*}_{li}|^{2}.
\end{equation}
This expression can be written in a more convenient form if we take into account
 the unitarity relation $\sum_{i}U_{l'i}~U^{*}_{li}=\delta_{l'l}$ and arbitrariness of a common phase.  We can have
\begin{equation}\label{Prob4}
\mathrm{P}(\nu_{l}\to \nu_{l'})=|\delta_{l'l}+\sum_{i\neq r}U_{l'i}(e^{-i2\Delta_{ri}}-1)U^{*}_{li}|^{2}=|\delta_{l'l}-2i\sum_{i\neq r}e^{-i\Delta_{ri}}U_{l'i}U^{*}_{li}\sin\Delta_{ri}|^{2}.
\end{equation}
Here
\begin{equation}\label{Prob4}
 \Delta_{ri}=\frac{\Delta m^{2}_{ri}L}{4E},
\end{equation}
where $L\simeq t$ is  neutrino source-detector distance and $r$ is an arbitrary fixed index.

The state of the right-handed flavor antineutrino $\bar\nu_{l}$  with momentum $\vec{p}$ is given by the relation
\begin{equation}\label{Prob5}
|\bar\nu_{l}\rangle=\sum^{3+n_{st}}_{i=1}U_{li}~|\nu_{i}\rangle_{R}\quad (l=e,\mu,\tau).
\end{equation}
Here $|\nu_{i}\rangle_{R}$ is the state of right-handed neutrino in the Majorana case (or antineutrino in the Dirac case)  with mass $m_{i}$, momentum $\vec{p}$ and energy $E_{i}$. For the probability of $\bar\nu_{l}\to \bar\nu_{l'}$ we find the following expression
\begin{equation}\label{Prob6}
\mathrm{P}(\bar\nu_{l}\to \bar\nu_{l'})=|\delta_{l'l}+\sum_{i\neq r}U^{*}_{l'i}(e^{-i2\Delta_{ri}}-1)U_{li}|^{2}=|\delta_{l'l}-2i\sum_{i\neq r}e^{-i\Delta_{ri}}U^{*}_{l'i}U_{li}\sin\Delta_{ri}|^{2}.
\end{equation}

It is evident that neutrino oscillations (described by the second term in (\ref{Prob4}) and (\ref{Prob6})) is an interference effect. Neutrino oscillations can take place if the inequality
\begin{equation}\label{Prob7}
|E_{i}-E_{r}|t\simeq \frac{|\Delta m_{ri}^{2}|L}{2E}\gtrsim 1,\quad i\neq r.
\end{equation}
is satisfied.  Notice that (\ref{Prob7}) is the time-energy uncertainty relation.\footnote{From (\ref{Prob7}) it follows that in  order to resolve a difference between $E_{i}$ and $E_{r}$ the time interval $t\gtrsim \frac{1}{|E_{i}-E_{r}|}$ is needed (see \cite{Bilenky:2011pk})}.

From (\ref{Prob4}) and (\ref{Prob6}) for $\nu_{l}\to \nu_{l'}$ ($\bar\nu_{l}\to \bar\nu_{l'}$) transition probability we find the following general expression
\begin{eqnarray}
&&\mathrm{P}(\nua{l}\to \nua{l'})
=\delta_{l'l }
-4\sum_{i\neq r}|U_{l i}|^{2}(\delta_{l' l } - |U_{l' i}|^{2})\sin^{2}\Delta_{ri}\nonumber\\
&&+8~\sum_{i>k;i,k\neq r}\mathrm{Re}~(U_{l' i}U^{*}_{l i}U^{*}_{l'
k}U_{l k})\cos(\Delta_{ri}-\Delta_{rk})\sin\Delta_{ri}\sin\Delta_{rk}\nonumber\\
&&\pm 8~\sum_{i>k;i,k\neq r}\mathrm{Im}~(U_{l' i}U^{*}_{l i}U^{*}_{l'
k}U_{l k})\sin(\Delta_{ri}-\Delta_{rk})\sin\Delta_{ri}\sin\Delta_{rk}
\label{Probabil}
\end{eqnarray}
\subsection{Three-Neutrino Oscillations}
Existing neutrino oscillation data are perfectly described under the minimal assumption of the three-neutrino mixing
\begin{equation}\label{3nu1}
 \nu_{lL}=\sum^{3}_{i=1} U_{li}\nu_{iL},\quad l=e,\mu,\tau.
\end{equation}
In the case of the three-neutrino mixing transition probabilities depend on six parameters: two mass-squared differences, three mixing angles and one phase. From analysis of the existing data it follows that one mass-squared difference   is about 30 times smaller than the other one.

The effect of small mass-squared difference was observed for the first time in the solar neutrino experiments. This is the reason why it is usually called  the ``solar mass-squared difference" $\Delta m^{2}_{S}$. It is determined by the following relation
\begin{equation}\label{3nu2}
\Delta m^{2}_{S}=\Delta m^{2}_{12}.
\end{equation}
From the solar neutrino data  follows that $\Delta m^{2}_{12}>0$. For the mass of the third neutrino $m_{3}$ there are two possibilities:
\begin{enumerate}
  \item  Normal ordering of neutrino masses (NO)
 \begin{equation}\label{3nu3}
    m_{3}>m_{2}>m_{1}.
 \end{equation}
\item Inverting ordering of neutrino masses (IO)
\begin{equation}\label{3nu4}
    m_{2}>m_{1}>m_{3}.
 \end{equation}
\end{enumerate}
For the first time the largest neutrino mass-squared difference was determined  from the data of atmospheric neutrino experiments. It is usually called  the ``atmospheric mass-squared difference". It is natural to determine the atmospheric mass-squared difference in such a way that it does not depend on (unknown) type of the  neutrino mass spectrum. One  possibility is the following\footnote{In the literature exist other definitions of the atmospheric mass-squared difference. The NuFit group \cite{Esteban:2016qun} uses the following definition:  $$\Delta m^{2}_{A}=\Delta m^{2}_{13} ~(NO),\quad
\Delta m^{2}_{A}=|\Delta m^{2}_{23}| ~(IO).$$ The Bari group \cite{Marrone:2016rmx} defines $\Delta m^{2}_{A}$ as follows $$\Delta m^{2}_{A}=\frac{1}{2}|\Delta m^{2}_{13}+\Delta m^{2}_{23}|$$}
\begin{equation}\label{Amass}
\Delta m^{2}_{A}=\Delta m^{2}_{23} ~(NO),\quad
\Delta m^{2}_{A}=|\Delta m^{2}_{13}|~(IO).
\end{equation}
In the case of the Dirac neutrinos $\nu_{i}$ the mixing matrix  is characterized by three mixing angles and one phase. In the standard parametrization it  has the form
\begin{eqnarray}
U^{\mathrm{D}}=\left(\begin{array}{ccc}c_{13}c_{12}&c_{13}s_{12}&s_{13}e^{-i\delta}\\
-c_{23}s_{12}-s_{23}c_{12}s_{13}e^{i\delta}&
c_{23}c_{12}-s_{23}s_{12}s_{13}e^{i\delta}&c_{13}s_{23}\\
s_{23}s_{12}-c_{23}c_{12}s_{13}e^{i\delta}&
-s_{23}c_{12}-c_{23}s_{12}s_{13}e^{i\delta}&c_{13}c_{23}
\end{array}\right).
\label{unitmixU1}
\end{eqnarray}
Here $\theta_{12}, \theta_{23}, \theta_{13}$ are Euler rotation angles, $c_{ik}=\cos\theta_{ik},~s_{ik}=\sin\theta_{ik}$. If the lepton sector $CP$ is conserved $U^{\mathrm{D}*}=U^{\mathrm{D}}$ and $\delta=0$. Thus, the phase $\delta$ characterizes  the $CP$ violation in the lepton sector.

If $\nu_{i}$ are Majorana particles, the mixing matrix is characterized by three angles and three phases and has the form
\begin{equation}\label{3Mjmix}
    U=U^{\mathrm{D}}~S(\alpha)
\end{equation}
where
\begin{eqnarray}\label{3Mjcol}
 S^{M}(\alpha)=\left(
\begin{array}{ccc}
e^{i\alpha_{1}}&0&0\\
0&e^{i\alpha_{2}}&0\\
0&0&1\\
\end{array}
\right).
\end{eqnarray}
Expressions for neutrino and antineutrino transition probabilities depend on the type of the neutrino mass spectrum. In the case of the normal ordering it is convenient to choose $r=2$. From the expression (\ref{Probabil}) for the $\nua{l}\to \nua{l'}$ transition probability we find the following expression
\begin{eqnarray}
&&\mathrm{P}^{\mathrm{NO}}(\nua{l}\to \nua{l'})
=\delta_{l' l }
-4|U_{l 3}|^{2}(\delta_{l' l} - |U_{l' 3}|^{2})\sin^{2}\Delta_{A}\nonumber\\&&-4|U_{l 1}|^{2}(\delta_{l' l} - |U_{l' 1}|^{2})\sin^{2}\Delta_{S}
-8~\mathrm{Re}~(U_{l' 3}U^{*}_{l 3}U^{*}_{l'
1}U_{l 1})\cos(\Delta_{A}+\Delta_{S})\sin\Delta_{A}\sin\Delta_{S}\nonumber\\
&&\mp 8~\mathrm{Im}~(U_{l' 3}U^{*}_{l 3}U^{*}_{l'
1}U_{l 1})\sin(\Delta_{A}+\Delta_{S})\sin\Delta_{A}\sin\Delta_{S}.
\label{Genexp5}
\end{eqnarray}
In the case of the inverted ordering we choose $r=1$. In this case from the expression (\ref{Probabil}) we obtain the following expression for the $\nua{l}\to \nua{l'}$ transition probability
\begin{eqnarray}
&&\mathrm{P}^{\mathrm{IO}}(\nua{l}\to \nua{l'})
=\delta_{l' l }
-4|U_{l 3}|^{2}(\delta_{l' l } - |U_{l' 3}|^{2})\sin^{2}\Delta_{A}\nonumber\\&&-4|U_{l 2}|^{2}(\delta_{l' l} - |U_{l' 2}|^{2})\sin^{2}\Delta_{S}
-8~\mathrm{Re}~(U_{l' 3}U^{*}_{l 3}U^{*}_{l'
2}U_{l 2})\cos(\Delta_{A}+\Delta_{S})\sin\Delta_{A}\sin\Delta_{S}\nonumber\\
&&\pm 8~\mathrm{Im}~(U_{l' 3}U^{*}_{l 3}U^{*}_{l'
2}U_{l 2})\sin(\Delta_{A}+\Delta_{S})\sin\Delta_{A}\sin\Delta_{S}.
\label{Genexp6}
\end{eqnarray}
Here
\begin{equation}\label{Delta}
\Delta_{A,S}=\frac{\Delta m^{2}_{A,S}L}{4E},
\end{equation}
It is obvious that
$P^{NO}(\nua{l}\to \nua{l'})$ and $P^{IO}(\nua{l}\to \nua{l'})$ differ by the change $U_{l 1}\to U_{l 2} $ ($U_{l' 1}\to U_{l' 2}$) and the change $(\mp)\to(\pm)$ in the last terms. From (\ref{Genexp5}) and (\ref{Genexp6})  expressions for all possible three-neutrino disappearance and appearance probabilities  can be easily obtained.

Values of the neutrino oscillation parameters obtained from the  global fit of all existing neutrino oscillation data \cite{Esteban:2018azc} are presented in the Table \ref{tab:1}.
\begin{table}
\caption{Values of neutrino oscillation parameters obtained from the
global fit of existing data}
\label{tab:1}
\begin{center}
\begin{tabular}{|c|c|c|}
  \hline  Parameter &  Normal Ordering& Inverted Ordering\\
\hline   $\sin^{2}\theta_{12}$& $0.310^{+0.013}_{-0.012}$& $0.310^{+0.013}_{-0.012}$
\\
\hline    $\sin^{2}\theta_{23}$& $0.580^{+0.017}_{-0.021}$& $ 0.584^{+0.016}_{-0.020}$
\\
\hline   $\sin^{2}\theta_{13}$ & $ 0.02241^{+0.00065}_{-0.00065}$&  $0.02264^{+0.00066}_{-0.00066}$
\\
\hline   $\delta $~(in $^{\circ}$) & $(215^{+40}_{-29})$& $ (284^{+27}_{-29})$
\\
\hline $\Delta m^{2}_{S}$& $(7.39^{+0.21}_{-0.20})\cdot 10^{-5}~\mathrm{eV}^{2}$&$(7.39^{+0.21}_{-0.20})\cdot 10^{-5}~\mathrm{eV}^{2}$\\
\hline $\Delta m^{2}_{A}$& $(2.525^{+0.033}_{-0.032})\cdot 10^{-3}~\mathrm{eV}^{2}$&$(2.512^{+0.034}_{-0.032})\cdot 10^{-3}~\mathrm{eV}^{2}$\\
\hline
\end{tabular}
\end{center}
\end{table}
It is expected that in the future neutrino oscillation experiments DUNE \cite{Abi:2018dnh}, JUNO \cite{Antonelli:2017uhq} and others
accuracy in the determination of the neutrino oscillation parameters will be significantly improved.

\section{On Neutrinoless Double $\beta$-Decay}
\subsection{Introduction}
If neutrinos with definite masses $\nu_{i}$ are Majorana particles in this case  lepton number violating neutrinoless double $\beta$-decay ($0\nu\beta\beta$-decay)
of $^{76}\mathrm{Ge}$,  $^{130}\mathrm{Te}$,  $^{136}\mathrm{Xe}$ and other even-even nuclei become possible. By different practical reasons (large targets (in future experiments about 1 ton or more), law backgrounds, high energy resolutions etc) the study of the process
\begin{equation}\label{bbdecay}
(A,Z) \to (A,Z+2) +e^{-}+ e^{-}
\end{equation}
 is the most sensitive method of the probe of the nature of $\nu_{i}$.

However,  the expected probability of the process (\ref{bbdecay}) is extremely small. This is connected with the fact that
\begin{itemize}
  \item The $0\nu\beta\beta$-decay is a process of the second order of the perturbation theory in the Fermi  constant $G_{F}$.
  \item   Because of the $V-A$ nature of the weak interaction the $0\nu\beta\beta$-decay   is due to the neutrino helicity flip.
 As a result, the matrix element of the decay is proportional  to the effective Majorana mass $m_{\beta\beta}=\sum_{i}U^{2}_{ei}m_{i}$. Smallness of the neutrino masses is a reason for the very strong suppression  of the probability of the $0\nu\beta\beta$-decay.
  \item Existing neutrino oscillation data favor the normal ordering of the neutrino mass spectrum. In the case of NO spectrum  $m_{\beta\beta}$ is much smaller than in the case of IO spectrum.
\item There could be additional  suppression of the probability of the $0\nu\beta\beta$-decay due to quenching of nuclear matrix elements (see \cite{Suhonen:2017krv}).

\end{itemize}
\section{Basics of the Phenomenological Theory of the $0\nu\beta\beta$-decay}
Assume that an even-even nucleus $(A,Z)$ has a mass $M_{A,Z}$ and that the mass of odd-odd nucleus  with the same atomic number  is larger than $M_{A,Z}$. In such a case the usual $\beta$-decay $(A,Z)\to (A,Z+1)+e^{-}+\bar\nu_{e}$ is forbidden. If, however, exist even-even
nucleus  $(A,Z+2)$ with mass smaller than $M_{A,Z}$, the nucleus
$(A,Z)$ can decay  into $(A,Z+2)$ with emission of two electrons via $L$-violating decay (\ref{bbdecay}) or $L$-conserving process
\begin{equation}\label{bbdecay1}
(A,Z)\to (A,Z+2)+e^{-}+e^{-}+\bar\nu_{e}+\bar\nu_{e}.
\end{equation}
In the Table \ref{tab:2}  a list of several $0\nu\beta\beta$ candidate nuclei are presented.
\begin{table}
\caption{$0\nu\beta\beta$ candidate nuclei. In the first column nuclei transitions are indicated; in the second column $Q$-values are shown; in the third column  abundances of the candidate nuclei are presented}
\label{tab:2}
\begin{center}
\begin{tabular}{|c|c|c|}
  \hline  transition &$T_{0}=Q_{\beta\beta}$($\rm{KeV}$)& Abundance (\%) \\
\hline   $^{76}\rm{Ge}\to ^{76}\rm{Se}$ & $2039.6\pm 0.9$ & 7.8
\\
\hline   $^{100}\rm{Mo}\to ^{100}\rm{Ru}$ & $3934\pm 6$ & 9.6
\\
\hline   $^{130}\rm{Te}\to ^{130}\rm{Xe}$ & $2533\pm 4$ & 34.5
\\
\hline   $^{136}\rm{Xe}\to ^{136}\rm{Ba}$ & $2479\pm 8$ & 8.9
\\
\hline   $^{150}\rm{Nd}\to ^{150}\rm{Sm}$ & $3367.1\pm 2.2$ & 5.6
\\
\hline   $^{82}\rm{Se}\to ^{82}\rm{Kr}$ & $2995\pm 6$ & 9,2
\\
\hline   $^{48}\rm{Ca}\to ^{48}\rm{Ti}$ & $4271\pm 4$ & 0.187
\\
\hline
\end{tabular}
\end{center}
\end{table}

We will consider here briefly the phenomenological theory of the neutrinoless double $\beta$-decay (see reviews \cite{Doi:1985,Bilenky:2014uka,Vergados:2016hso}). The SM effective Hamiltonian of the $\beta$-decay is given by the expression
\begin{equation}\label{effham}
{\mathcal{H}}_{I}(x)= \frac{G_F}{\sqrt{2}} 2~\bar e_{L}(x)
\gamma_{\alpha}~ \nu_{eL}(x)~j^{\alpha}(x) + \mathrm{h.c.}
\end{equation}
where $j^{\alpha}(x)$ is the hadronic $\Delta S=0$ charged current.

Neutrino oscillation data are in perfect agreement with the three-neutrino mixing
\begin{equation}\label{3numix}
\nu_{eL}(x) = \sum^{3}_{i=k} U_{ek} \nu_{kL}(x),
\end{equation}
where $U_{ek}$ are elements of the first row of the PMNS mixing matrix.
We will assume that  {\em $\nu_{k}(x)$  is the  field of the Majorana neutrino with mass $m_{k}$}.

The process (\ref{bbdecay}) is the second order in  $G_{F}$ process with virtual neutrino. From (\ref{effham}) and (\ref{3numix}) for the matrix element of the
$0\nu\beta\beta$-decay we find the following expression
\begin{eqnarray}\label{Smatelem}
&&\langle f|S^{(2)}|i
\rangle=4\frac{(-i)^{2}}{2~!}~\left (\frac{G_F}{\sqrt{2}}\right )^{2}N_{p_1}N_{p_2}
 \int\bar
u_{L}(p_1)e^{ip_{1}x_{1}}\gamma_{\alpha}~\langle 0|T(\nu_{eL}(x_{1})~\nu^{T}_{eL}(x_{2})|0\rangle
\nonumber\\
&&\times \gamma^{T}_{\beta}~\bar
u^{T}_{L}(p_2)e^{ip_{2}x_{2}}\langle
N_{f}|T(J^{\alpha}(x_{1})J^{\beta}(x_{2}))|N_{i}
\rangle~d^{4}x_{1}d^{4}x_{2}-(p_{1}\rightleftarrows p_{2}).
\end{eqnarray}
Here $p_{1}$ and $p_{2}$ are electron momenta, $J^{\alpha}(x)$
is the weak charged current in the Heisenberg representation, $N_{i}$ and $N_{f}$ are the states of the initial and the final nuclei
with 4-momenta $P_{i}=(E_{i}, \vec{p_{i}})$ and  $P_{f}=(E_{f}, \vec{p_{f}})$, respectively, and $N_{p}=\frac{1}{(2\pi)^{3/2}\sqrt{2p^{0}}}$ is the standard
normalization factor.

Taking into account  the Majorana condition $\nu_{k}=\nu^{c}_{k}=C\bar\nu^{T}_{k}$ for the neutrino propagator we have
\begin{eqnarray}\label{nuprop}
&&\langle 0|T(\nu_{eL}(x_{1})~\nu^{T}_{eL}(x_{2})|0\rangle=
-\sum_{k}U^{2}_{ek}\frac{1-\gamma_{5}}{2}\langle 0|T(\nu_{k}(x_{1})~\bar\nu_{k}(x_{2})|0\rangle\frac{1-\gamma_{5}}{2}C
\nonumber\\&&=-\sum_{k}U^{2}_{ek}m_{k}~\frac{i}{(2\pi)^{4}}
\int \frac{e^{-iq~(x_{1}-x_{2})}}{q^{2}-m^{2}_{k}} d^{4}q~\frac{1-\gamma_{5}}{2}C.
\end{eqnarray}
From (\ref{Smatelem}) and (\ref{nuprop}) for the matrix element of
$0\nu\beta\beta$-decay we obtain the following expression
\begin{eqnarray}\label{Smatelem1}
\langle f|S^{(2)}|i
\rangle&=&-4 \left (\frac{G_F}{\sqrt{2}}\right )^{2}N_{p_1}N_{p_2}
\sum_{k}U^{2}_{ek}m_{k} \int\bar
u_{L}(p_1)e^{ip_{1}x_{1}}\gamma_{\alpha}\left(\frac{i}{(2\pi)^{4}}
\int \frac{e^{-iq~(x_{1}-x_{2})}}
{ q^{2}-m^{2}_{k}}d^{4}q\right)\nonumber\\
&&\times \gamma_{\beta}\frac{1+\gamma_{5}}{2}C~\bar
u^{T}_{L}(p_2)e^{ip_{2}x_{2}} \langle
N_{f}|T(J^{\alpha}(x_{1})J^{\beta}(x_{2}))|N_{i}
\rangle~d^{4}x_{1}d^{4}x_{2}.
\end{eqnarray}
In this expression we can perform integration over $x^{0}_{1}$, $x^{0}_{2}$ and $q^{0}$. We find
\begin{eqnarray}\label{Smatelem2}
&&\langle f|S^{(2)}|i
\rangle=2i~\left(\frac{G_F}{\sqrt{2}}\right )^{2}N_{p_1}N_{p_2}
 \bar
u(p_1)\gamma_{\alpha}\gamma_{\beta}(1+\gamma_{5})C \bar
u^{T}(p_2)\int
d^{3}x_{1}d^{3}x_{1}e^{-i\vec{p}_{1}\vec{x}_{1}-i\vec{p}_{2}\vec{x}_{2}}
 \nonumber\\
&&\times\sum_{k}U^{2}_{ek}m_{k}\frac{1}{(2\pi)^{3}}
\int\frac{e^{i\vec{q}~(\vec{x}_{1}-\vec{x}_{2})}} {
q_{k}^{0}}d^{3}q~[\sum_{n} \frac{\langle
N_{f}|J^{\alpha}(\vec{x}_{1})|N_{n}\rangle\langle N_{n}|
J^{\beta}(\vec{x}_{2}))|N_{i}\rangle }{E_{n}+p^{0}_{2}+q^{0}_{k}-E_{i}-i\epsilon}
\nonumber\\ &&+\sum_{n}\frac{ \langle
N_{f}|J^{\beta}(\vec{x}_{2})|N_{n}\rangle\langle N_{n}|
J^{\alpha}(\vec{x}_{1}))|N_{i}\rangle}{E_{n}+p^{0}_{1}+q^{0}_{k}-E_{i}-i\epsilon}
]~2\pi\delta(E_{f}+p^{0}_{1}+p^{0}_{2}-E_{i}),
\end{eqnarray}
where $q^{0}_{k}=\sqrt{q^{2}+m_{k}^{2}}$ ($q=|\vec{q}|$) and $|N_{n}\rangle$ is the vector of the state of the intermediate nucleus with 4-momentum $P_{n}=(E_{n},\vec{p}_{n}$). In (\ref{Smatelem2}) the sum over the total system of the states $|N_{n}\rangle$ is assumed.

The equation (\ref{Smatelem2}) is an exact expression for the matrix element of the $0\nu\beta\beta$-decay in the second order of the perturbation theory. It is obvious that
\begin{itemize}
  \item Small neutrino masses can be neglected in the expression for the virtual neutrino energy $q^{0}_{k}$. In fact, the average neutrino momentum is given by  $\bar{q}\simeq \frac{1}{r}$, where $r\simeq 10^{-13}$ cm is the average distance between nucleons in a nucleus. We have
 $\bar{q}\simeq 100~\mathrm{MeV}\gg m_{k}$.

\item We have $|\vec{p}_{i}\cdot
\vec{x}_{i}|\leq p_{i}R$~($i=1,2$), where $R\simeq 1.2\cdot10^{-13}~A^{1/3}~\mathrm{cm}$ is the radius of a nucleus. Taking into account that $p_{i}\lesssim \mathrm{MeV}$ we find that $|\vec{p_{i}}\cdot\vec{x_{i}}|\ll 1$ and $e^{-i\vec{p}_{1}\vec{x}_{1}-i\vec{p}_{2}\vec{x}_{2}}\simeq 1$.

\end{itemize}
The matrix element of the $0\nu\beta\beta$-decay takes the form
\begin{eqnarray}\label{Smatelem3}
&&\langle f|S^{(2)}|i
\rangle\simeq m_{\beta\beta}~2i~\left(\frac{G_F}{\sqrt{2}}\right )^{2}N_{p_1}N_{p_2}
 \bar
u(p_1)\gamma_{\alpha}\gamma_{\beta}(1+\gamma_{5})C \bar
u^{T}(p_2)
 \nonumber\\
&&\times\int
d^{3}x_{1}d^{3}x_{1}\frac{1}{(2\pi)^{3}}
\int\frac{e^{i\vec{q}~(\vec{x}_{1}-\vec{x}_{2})}} {
q}d^{3}q~[\sum_{n} \frac{\langle
N_{f}|J^{\alpha}(\vec{x_{1}})|N_{n}\rangle\langle N_{n}|
J^{\beta}(\vec{x_{2}}))|N_{i}\rangle }{E_{n}+p^{0}_{2}+q
-E_{i}-i\epsilon}
\nonumber\\ &&+\sum_{n}\frac{ \langle
N_{f}|J^{\beta}(\vec{x_{2}})|N_{n}\rangle\langle N_{n}|
J^{\alpha}(\vec{x_{1}}))|N_{i}\rangle}{E_{n}+p^{0}_{1}+q
-E_{i}-i\epsilon}
]~2\pi\delta(E_{f}+p^{0}_{1}+p^{0}_{2}-E_{i}),
\end{eqnarray}
where
\begin{equation}\label{eMjmass}
 |m_{\beta\beta}|=|\sum^{3}_{k=1}U^{2}_{ek}~m_{k}|
\end{equation}
is {\em the effective Majorana mass.}

Calculation of a nuclear part of the matrix element of the $0\nu\beta\beta$-decay is a complicated many-body nuclear problem. Different approximate  methods are used in such calculations. Discussion of such calculations is out of the scope of this review. Let us stress that at the moment the results of different calculations differ by 2-3 times~(see reviews \cite{Vergados:2016hso,Engel:2016xgb}).

From
(\ref{Smatelem3}) it follows that the half-life of the
$0\nu\beta\beta$-decay has the following general form
\begin{equation}\label{totrate}
\frac{1}{T^{0\nu}_{1/2}}=|m_{\beta\beta}|^{2}~|M^{0\nu}|^{2}~
G^{0\nu}(Q,Z).
\end{equation}
Here $M^{0\nu}$ is the nuclear matrix element (NME) and $G^{0\nu}(Q,Z)$ is the known phase-space factor.

The effective Majorana mass $|m_{\beta\beta}|$ is determined by neutrino masses, neutrino mixing angles and Majorana $CP$ phases.

In the case of NO of neutrino masses  the  masses $m_{2,3}$ are
connected with the lightest mass $m_{1}$ and two neutrino mass-squared differences by the relations
\begin{equation}\label{norspec1}
m_{2}=\sqrt{m^{2}_{1}+\Delta m^{2}_{S}},~~
m_{3}=\sqrt{m^{2}_{1}+\Delta m^{2}_{S}+\Delta m^{2}_{A}}
\end{equation}
In the case of IO of neutrino masses we have
\begin{equation}\label{invspec1}
m_{1}=\sqrt{m^{2}_{3}+\Delta m^{2}_{A}},\quad
m_{2}=\sqrt{m^{2}_{3}+\Delta m^{2}_{A}+\Delta m^{2}_{S}}
\end{equation}

The following  neutrino mass spectra are of special interest.
\begin{center}
  {\em Hierarchy of the neutrino masses}
\end{center}
\begin{equation}\label{hierar}
m_{1} \ll m_{2} \ll m_{3}.
\end{equation}
In this case we have
\begin{equation}\label{hierar1}
 m_{2}\simeq \sqrt{ \Delta
m^{2}_{S}},\quad m_{3}\simeq  \sqrt{ \Delta m^{2}_{A}}\quad
m_{1} \ll \sqrt{\Delta m^{2}_{S}}.
\end{equation}
Thus, in the case of the neutrino mass hierarchy  masses $m_{2}$ and $m_{3}$ are determined by the solar and atmospheric mass-squared differences, respectively, and lightest mass $m_{1}$ is very small. Neglecting its contribution  and using the standard parametrization of the PMNS mixing matrix  for the effective Majorana mass we have
\begin{equation}\label{hierar2}
|m_{\beta\beta}|\simeq\left |\,\cos^{2}\theta_{13}
 \sin^{2} \theta_{12}\, \sqrt{\Delta m^{2}_{S}} + e^{2i\,\alpha}
 \sin^{2} \theta_{13}\, \sqrt{\Delta m^{2}_{A}}\,\right |~,
\end{equation}
where $\alpha$ is the Majorana phase difference.

The first term in Eq.(\ref{hierar2}) is small because of the
smallness of $\Delta m^{2}_{S}$. The contribution of the ``large''
$\Delta m^{2}_{A}$ to $|m_{\beta\beta}|$ is suppressed by
the smallness of $\sin^{2} \theta_{13} $.
Using the best-fit values of the parameters we have
\begin{equation}\label{globfit1}
 \cos^{2}\theta_{13}\sin^{2} \theta_{12}\, \sqrt{\Delta
m^{2}_{S}}\simeq 3\cdot 10^{-3}~\mathrm{eV},\quad
\sin^{2} \theta_{13}\,
\sqrt{\Delta m^{2}_{A}}=  1\cdot 10^{-3}\rm{eV}.
\end{equation}
Thus, the absolute values of the first and second terms in (\ref{hierar2}) are of the same order of magnitude. Taking into account errors and all possible values of the phase $\alpha$ we have
\begin{equation}\label{globfit2}
|m_{\beta\beta}|\leq 4\cdot 10^{-3}\rm{eV}.
\end{equation}
The upper bound (\ref{globfit2}) is significantly smaller than the expected sensitivity of the future experiments on the search for the $0\nu\beta\beta$-decay (see later).

\begin{center}
  {\em Inverted hierarchy of the neutrino masses}
\end{center}
\begin{equation}\label{invhierar}
m_{3} \ll m_{1} < m_{2}.
\end{equation}
In this case from (\ref{invspec1})  we have
\begin{equation}\label{invhierar1}
m_{1}\simeq \sqrt{\Delta m^{2}_{A}},~~m_{2}\simeq\sqrt{ \Delta m^{2}_{A}}~(1+\frac{\Delta m^{2}_{S}}{2\, \Delta m^{2}_{A}})\simeq\sqrt{ \Delta m^{2}_{A}},~~m_{3}\ll\sqrt{
\Delta m^{2}_{A}}.
\end{equation}
In the expression for
$|m_{\beta\beta}|$
the lightest mass $m_{3}$ is multiplied by the small parameter
$\sin^{2}\theta_{13}$. Neglecting the contribution of this term, we find
\begin{equation}\label{invhierar2}
|m_{\beta\beta}|\simeq \sqrt{ \Delta m^{2}_{A}}\cos^{2}\theta_{13}~ (1-\sin^{2}
2\,\theta_{12}\,\sin^{2}\alpha)^{\frac{1}{2}},
\end{equation}
where $\alpha$ is  the Majorana phase difference, the only unknown parameter in the expression (\ref{invhierar2}). From (\ref{invhierar2}) we find the following upper and lower bounds for the effective Majorana mass
\begin{equation}\label{invhierar3}
\cos^{2}\theta_{13}~\cos  2\,\theta_{12} \,\sqrt{ \Delta m^{2}_{A} } \leq
|m_{\beta\beta}| \leq\cos^{2}\theta_{13}~\sqrt{ \Delta m^{2}_{A}}.
\end{equation}
Notice that the upper and lower bounds of this inequality
 correspond to the $CP$-invariance in the lepton sector. From  (\ref{invhierar3}) we find the following range for the effective Majorana mass
\begin{equation}\label{invhierar4}
2\cdot 10^{-2}~\mathrm{eV}\leq  |m_{\beta\beta}|\leq 5\cdot 10^{-2}~\mathrm{eV}
\end{equation}
Future experiments on the search for $0\nu\beta\beta$-decay will probe the region
(\ref{invhierar4}).

\subsection{Results of Experiments on the Search for $0\nu\beta\beta$-decay}
In the Table \ref{tab:3} results of the recent experiments on search for
$0\nu\beta\beta$-decay are presented. The process was not observed. In the forth column of the Table \ref{tab:3} the 90\% CL lower bounds for the half-live of the decay of different elements are given. Taking into account values of nuclear matrix elements, obtained by different authors, and presented in the second column of the Table we give in the fifth column upper bounds of the effective Majorana mass. As it is seen from the Table \ref{tab:3} the present experiments still do not reach the inverted hierarchy region.
\begin{table}
\caption{Lower limits of half-lives $T^{0\nu}_{1/2}$ and upper limits of the effective Majorana mass  $|m_{\beta\beta}|$ obtained in recent experiments on the search for the $0\nu\beta\beta$-decay}
\label{tab:3}
\begin{center}
\begin{tabular}{|c|c|c|c|c}
  \hline experiment& nucleus& NME& $T^{0\nu}_{1/2}$($10^{25}$yr)&  $|m_{\beta\beta}|$ (eV) \\
\hline Gerda \cite{Agostini:2018tnm}&$^{76}$ Ge & 2.8-6.1 & 8.0& (0.12-0.26) \\
  \hline Majorana \cite{Alvis:2019sil}&$^{76}$ Ge& 2.8-6.1 & 2.7& (0.20-0.43) \\
  \hline KamLAND-Zen \cite{KamLAND-Zen:2016pfg} & $^{136}$ Xe&1.6-4.8& 10.7& (0.05-0.16) \\
  \hline EXO \cite{Albert:2017owj} & $^{136}$ Xe&1.6-4.8   & 1.8& (0.15-0.40)   \\
\hline CUORE \cite{Alduino:2017ehq}&$^{130}$ Te&1.4-6.4    &1.5& (0.11-0.50)\\
\hline
\end{tabular}
\end{center}
\end{table}
Many new experiments on the search for the $0\nu\beta\beta$-decay are in preparation at present. In these future experiments the inverted hierarchy region and, possibly, part of the normal hierarchy region will be probed. Projected sensitivities of future experiments are presented in the Table \ref{tab:4} (see \cite{Giuliani:2018}).

\begin{table}
\caption{Projected sensitivities  of future experiments on the search for the $0\nu\beta\beta$-decay (lower limits of half-lives and upper limits of the effective Majorana mass)}
\label{tab:4}
\begin{center}
\begin{tabular}{|c|c|c|c|c}
  \hline experiment& nucleus& $T^{0\nu}_{1/2}$&  $|m_{\beta\beta}|$  \\
\hline nEXO &$^{136}$ Xe  & $9.2\cdot 10^{27}$yr& (5-20)~$\mathrm{meV}$ \\
  \hline NEXT100&$^{136}$ Xe& $ 9.8\cdot 10^{25}$yr&(46-170)~$\mathrm{meV}$ \\
  \hline KamLAND-Zen800  & $^{136}$ Xe& $4.6\cdot 10^{26}$yr&(25-80)$~\mathrm{meV}$  \\
  \hline LEGEND  & $^{76}$ Ge& $1\cdot 10^{27}$yr& (35-75)~$\mathrm{meV}$   \\
\hline CUORE &$^{130}$ Te&$1\cdot 10^{26}$yr& (50-190)$~\mathrm{meV}$\\
\hline
\end{tabular}
\end{center}
\end{table}

\section{On Sterile Neutrinos Search}

\subsection{Indications in Favor of Sterile Neutrinos}

If indications in favor of normal ordering of the  neutrino spectrum, obtained in accelerator and atmospheric neutrino oscillation experiments, will be confirmed by future experiments, observation of the $0\nu\beta\beta$-decay could require many years and, possibly, new technologies. On the other side, the problem of sterile neutrinos, apparently, will be resolved in the nearest years.

The following indications in favor of the sterile neutrinos were obtained.

{\bf LSND}. In the short baseline LSND accelerator beam-dump experiment \cite{Aguilar:2001ty} neutrinos were produced in $\pi^{+}\to \mu^{+}+\nu_{\mu}$ and $\mu^{+}\to e^{+}+\nu_{e}+ \bar\nu_{\mu}$ decays at rest.
At a distance about 30 m from the neutrino source in the LSND detector
$\bar\nu_{e}+p\to e^{+}+n$ events were searched for. An excess $87.9\pm 22.4\pm 6.0$ events were observed. These events could be explained by $\bar\nu_{\mu}\to \bar\nu_{e}$ oscillations with the average  oscillation probability equal to $(0.264\pm0.067\pm0.045)\%$. From analysis of the data it was found that the region
$0.2\leq \Delta m^{2}\leq 10 ~\mathrm{eV}^{2}$ was allowed. The best-fit values of  the neutrino oscillation parameters were $\sin^{2}2\theta=0.003,~~\Delta m^{2}=1.2~\mathrm{eV}^{2}$.

{\bf MiniBooNE}. The MiniBooNE experiment  at the Fermilab was performed with the aim to check  the LSND result.  The average value of the parameter $\frac{L}{E}$ in this experiment was approximately the same as in the LSND experiment ($\simeq 1\mathrm{m}/\mathrm{MeV}$). Recently the result of the running of the experiment during 15 years was published \cite{Aguilar-Arevalo:2018gpe}.
In the  $\nu_{\mu}$ and $\bar\nu_{\mu}$ runs $460.5\pm 99.0$
$\nu_{e}$ and $\bar\nu_{e}$ events in excess to expectation were found in the energy interval (200-1250)~MeV. This result can be explained by neutrino oscillations and is compatible with the LSND result.

{\bf Reactor neutrino anomaly}. In 2011 reactor antineutrino spectrum was recalculated. As a result of new calculations the mean flux of reactor $\bar\nu_{e}$'s increased by 3.5\% and the ratio of measured and expected $\bar\nu_{e}$ events in all performed reactor short baseline neutrino experiments became equal to $0.943\pm0.023$ (previous ratio was equal to $0.976\pm0.024$). Thus, with the new flux results of old reactor neutrino experiments could be considered as an indication in favor of neutrino oscillations. From analysis of the data of reactor neutrino experiments it was found \cite{Mention:2011rk} that $\Delta m^{2}>1.5~\mathrm{eV}^{2}$, $\sin^{2}2\theta=0.14\pm 0.08$.

{\bf Galium neutrino anomaly}. In short baseline calibration neutrino experiments performed by the GALLEX and SAGE collaborations the measured ratios $R$ of detected and expected neutrino events was less than one. Neutrinos from radioactive sources were detected via observation of the process $\nu_{e}+^{71}\mathrm{Ga} \to e^{-}+^{71}\mathrm{Ge}$. In the GALLEX experiment ($^{51}\mathrm{Cr}$ source) it was found $R= 0.812^{+0.10}_{-0.11}$. In the SAGE experiment  ($^{37}\mathrm{Ar}$ source) it was obtained $R= 0.791^{+0.084}_{-0.078}$. These data can be explained by neutrino oscillations \cite{Giunti:2010zu}. From analysis of the data it was found
$\Delta m^{2}>0.35~\mathrm{eV}^{2}$, $\sin^{2}2\theta>0.07 $ at 99\% CL.

\subsection{3+1 Mixing Scheme. Appearance-disappearance tension}
Existing short baseline (SBL) neutrino oscillation data can be explained if we assume that in addition to three light neutrinos  exist a fourth neutrino $\nu_{4}$
with mass $m_{4}$  of the order of one eV. Average values of $\frac{L}{E}$ in the SBL experiments can be determined from the condition
\begin{equation}\label{SBL}
 \Delta _{14}=\frac{\Delta m_{14}^{2}L}{4E}\simeq 1,\quad \Delta m_{14}^{2}= m_{4}^{2}- m_{1}^{2}
\end{equation}
 For such values of the parameter $\frac{L}{E}$ we have $\Delta _{S}\ll\Delta _{A}\ll 1$. Thus, contributions of the atmospheric and solar mass-squared differences to the  SBL neutrino transition probabilities  can be safely neglected.

  From the general expression (\ref{Probabil})  we find
\begin{equation}\label{SBL2}
 P^{\mathrm{SBL}}(\nua{l}\to \nua{l'})\simeq \delta_{l'l}-4|U_{l4}|^{2}(\delta_{l'l}-
|U_{l'4}|^{2})\sin^{2}\Delta_{14}.
\end{equation}
Let us consider SBL transitions
\begin{equation}\label{SBL3}
\nua{\mu}\to \nua{e},~~\nua{e}\to \nua{e},~~\nua{\mu}\to \nua{\mu}
\end{equation}
 which are investigating in different experiments.

From (\ref{SBL2}) we find
\begin{eqnarray}\label{SBL5}
P^{\mathrm{SBL}}(\nua{e}\to \nua{e}) &=&1-4|U_{e4}|^{2}(1-
|U_{e4}|^{2})~\sin^{2}\Delta_{14
}\nonumber\\P^{\mathrm{SBL}}(\nua{\mu}\to \nua{\mu}) &=&1-4|U_{\mu 4}|^{2}(1-
|U_{\mu4}|^{2})~\sin^{2}\Delta_{14}.
\end{eqnarray}
The $\nua{\mu}\to \nua{e}$ appearance probability is given by
\begin{equation}\label{SBL6}
P^{\mathrm{SBL}}(\nua{\mu}\to \nua{e}) =\sin^{2}2\theta_{e\mu }\sin^{2}\Delta_{14}    \end{equation}
where
\begin{equation}\label{SBL7}
\sin^{2}2\theta_{e\mu }=4|U_{e 4}|^{2}|U_{\mu 4}|^{2}.
\end{equation}
The equation (\ref{SBL7}) is an important relation between SBL transition amplitudes \cite{Bilenky:1996rw,Okada:1996kw}. This relation follows from the fact that
amplitudes of  transitions (\ref{SBL3}) are determined by  parameters $|U_{e4}|^{2}$ and $|U_{\mu4}|^{2}$.

From (\ref{SBL7}) it follows that if results of LSND, MiniBooNE, SBL reactor and source experiments can be interpreted as neutrino oscillations, then from
data of these experiments  the oscillation amplitude in SBL $\nua{\mu}\to \nua{\mu}$ transition can be predicted. From analysis of the LSND and other data it was found that $|U_{\mu4}|^{2}\simeq 10^{-1}$.

Effects of neutrino oscillations was not observed in MINOS/MINOS+\cite{Adamson:2017uda} and other experiments on the search for $\nua{\mu}\to \nua{\mu}$ transition. From analysis of the existing data it was found that $|U_{\mu4}|^{2}< 10^{-2}$ in the region $2\cdot 10^{-1}< \Delta m^{2}_{14}<10~\mathrm{eV}^{2}$ (see, for example, \cite{Dentler:2018sju,Gariazzo:2017fdh})

This is a clear contradiction  to the interpretation of LSND and other results as neutrino oscillations. Definitely  {\em new more precise  experiments are urgently needed.}

\subsection{New  Experiments on the Search for Light  Sterile Neutrinos}
Many new short baseline neutrino experiments  on the search for sterile neutrinos with masses of the order of  eV are going on and in preparation (see
\cite{Giunti:2019aiy,Dentler:2018sju}). We will briefly discuss  new SBL reactor neutrino experiments (for a critical review see \cite{Danilov:2018dme}).

The {\bf DANSS} detector is a 1 $\mathrm{m}^{3}$ plastic scintillator. The experiment is performing at 3.1 $\mathrm{GW_{th}}$ power reactor in Russia. The detector is installed on a movable platform. Measurements are performed at the distances 10.7 m, 11.7 m and 12.7 m from the reactor center. This allows to carry out an analysis of the data which does not depend on the reactor antineutrino spectrum and detector efficiency. However, the energy resolution in the DANSS experiment is relatively poor ($\frac{\sigma}{E}\simeq 35 \%$ at 1 MeV). About 5000 $\bar\nu_{e}$ events per day are detected. Analysis of obtained data allows to exclude significant part of the $\Delta m^{2}_{14}, |U_{e4}|^{2}$ region allowed by previous $\nua{e}$ disappearance experiments. In particular, the previous best-fit point is excluded at more than 5 $\sigma$.

The {\bf NEOS} detector is 0.8 tons Gd-doped liquid scintillator. The experiment is performing  at 2.8 $\mathrm{GW_{th}}$ power reactor in Korea. The distance between detector and the reactor core is fixed at $23.7\pm 0.3$ m. The energy resolution in the experiment is very good: $\frac{\sigma}{E}\simeq 5 \%$ at 1 MeV. About 2000 neutrino events per day are detected. In order to avoid dependence on reactor neutrino spectrum the measured positron spectrum is compared with the spectrum measured in the Daya Bay experiment. A large region in the plane of oscillation parameters is excluded by the NEOS experiment.

The {\bf PROSPECT} detector is  4 ton $^{6}\mathrm{Li}$-loaded, segmented liquid scintillator. It is located at the high power compact reactor center (85 MW) of the Oak Ridge National Laboratory. High-energy resolution is achieved ($\frac{\sigma}{E}\simeq 4.5 \%$ at 1 MeV). The distance between reactor core and detector is 6.7 m. During 33 days $25461\pm 283$ antineutrino events were detected. No indications in favor of neutrino oscillations were found. The Reactor Antineutrino Anomaly best-fit point is excluded at 2.2 $\sigma$.

The {\bf STEREO} detector consist of six optically separated cells filled with
Gd-loaded  liquid scintillator. The total volume is about 2 $\mathrm{m}^{3}$.  The experiment is performing at 58 MW High Flux Reactor of the Institute Laue-Langevin (France).
The cell distances from the reactor core varies from 9.4 m to 11.1 m. Ratios of positron spectra in different cells  to the spectrum in the first cell is analyzed. The result of the 66 days of reactor on and 138 days of reactor off is compatible with no oscillations. A significant part of the previously allowed area in the plane of the oscillation parameters is excluded.

The {\bf NEUTRINO-4} detector consist of 50 sections filled with Gd-loaded liquid scintillator. The total volume of the detector  is 1.8 $\mathrm{m}^{3}$. The experiment is carried out  at a compact research reactor (100 MW) at Dmitrovgrad (Russia). The detector is installed on a movable platform which allow to make measurements at the distances from 6 m to 12 m. There is practically no overburden and a signal to background ratio is 0.54. Energy resolution is $ 16 \%$ at 1 MeV. Indications in favor of neutrino oscillations were found in the experiment. However, the best-fit point ($\Delta m^{2}_{14}= 7.22 ~\mathrm{eV}^{2},~~ \sin^{2}2\theta_{ee}=0.35$) is in a tension with the PROSPECT  data.

There are many other neutrino oscillation experiments on the search for sterile neutrinos (see \cite{Boser:2019rta}). Some ambitious experiments are in preparation. We would like to mention the SNO Program at the Fermilab \cite{Antonello:2015lea}). In this experiment three Liquid Argon Time Projection Detectors will be used: SBND (112 tons, 110 m  from the source), MicroBooNE (87 tons, 470 m), ICARUS (475 tons, 600 m). Important feature of this experiment will be {\em simultaneous measurement of $\nu_e$ appearance and
$\nu_\mu$ disappearance} in  MicroBooNE and ICARUS detectors. The sensitivity of the experiment will allow to exclude the LSND and MiniBooNE allowed regions and global best-fit points  at least at the $5 \sigma$ level.There is a good chance than {\em sterile neutrinos anomaly} will be resolved in the nearest years.

Let us stress again that in the most economical approach to the origin of neutrino masses, based on massless neutrinos in the Standard Model, on dimesion five $SU_{L}(2)\times U_{Y}(1)$ invariant effective Lagrangian in which only SM fields enter and on violation of the Lepton number in a beyond the SM theory, the number of  massive Majorana neutrinos is equal to the number of flavor neutrinos (three) and {\em there is no room for sterile neutrinos}. If light sterile neutrinos will be discovered in future short baseline and other neutrino experiments in this case  our ideas about the origin of small neutrino masses will be completely changed.

Let us notice that exist  several exotic models of sterile neutrinos in the literature. Because of arbitrariness of the Yukawa couplings the sterile neutrinos can be introduced  in the framework of the seesaw mechanism (see \cite{deGouvea:2005er}). Other models include extra dimensions, broken $L_{e}-L_{\mu}-L_{tau}$ symmetry, mirror particles etc (see, for example, \cite{Abazajian:2012ys}  and references therein).

\section{Conclusion}
The Standard Model is a great achievement of the physics of the XX century. All predictions of the Standard Model (Neutral Currents, $W^{\pm}$ and $Z^{0}$ bosons, $t$-quark, $\nu_{\tau}$ and many others) were perfectly confirmed by experiments. The  discovery of the Higgs boson at LHC was an impressive confirmation of the mechanism of the spontaneous symmetry breaking.

The Standard Model is  based on
the  local gauge invariance, unification of the weak and electromagnetic interactions, spontaneous symmetry breaking. In the framework of these principles in the Standard Model minimal possibilities are realized. Impressive agreement of the Standard Model with experiment signifies that  {\em the nature chooses the simplest and most economical possibilities.}

In this review we discuss the problem of neutrino masses. Neutrinos are unique particles: they are the only fundamental fermions which can be Dirac or Majorana particles. If we assume that the Standard Model includes right-handed neutrino fields, $SU(2)\times U(1)$ singlets, and neutrino masses and mixing can be generated by the Standard Higgs Mechanism (like masses of leptons and quarks) the total lepton number $L$ will be conserved and neutrinos are Dirac particles. Because of the smallness of neutrino masses with respect to the SM lepton and quark masses this possibility  is extremely unlikely (in spite it is not excluded experimentally). Thus it is very plausible that neutrino masses and mixing are generated by a beyond the Standard Model mechanism.

There are many beyond the SM models of the generation of small neutrino masses and neutrino mixing. Apparently,  the  lepton number violating, dimension five  Weinberg effective Lagrangian provides the most simple and economical possibility. It is based  on the assumption that SM neutrinos are massless particles (there are no right-handed neutrinos in the SM). The Weinberg Lagrangian generates Majorana neutrino mass term with Majorana neutrino masses which are suppressed with respect to the SM masses  of leptons and quarks by a factor which is given the ratio of the electroweak scale $v\simeq 246$ GeV and a (apparently, large but unknown) scale of a new, beyond the SM physics $\Lambda$.

The Weinberg Lagrangian can be generated in the tree approximation by different beyond the SM interactions: interaction of Higgs-lepton pair with heavy Majorana scalar or triplet lepton
(type I or type III seesaw) or interaction of heavy scalar triplet with Higgs and lepton pairs (type II  seesaw). It is a very attractive feature of these models that they could solve one of the greatest cosmological problem, the problem of the barion asymmetry of the Universe.

There are many other models in which the Weinberg effective Lagrangian can be generated at a loop level.  Such models lead to the Majorana neutrino mass term. A general feature of all these models is the absence of the right-handed neutrino fields in the SM. It is obvious that the values of small neutrino masses can not be predicted without additional information about a new beyond the SM physics.

However, general features of the  Majorana mass term which can be tested in the present-day neutrino experiments are the following:
\begin{itemize}
  \item Neutrinos with definite masses $\nu_{i}$ are Majorana particles
  \item The number of neutrino with definite masses is equal to the number of flavor neutrinos (three)
  \end{itemize}
As it is very well known, the nature of  $\nu_{i}$ (Dirac or Majorana?) can be probed in experiments on the search for  the neutrinoless double $\beta$-decay of
$^{76}\mathrm{Ge}$ and other even-even nuclei. We briefly reviewed here the results of these experiments and future perspectives.

If the number of light neutrinos with definite masses is more than three in this case transitions of flavor neutrinos into sterile states will take place. In this review we briefly discussed the present-day situation with the search for sterile neutrinos.

Let us remind that at the  time of the LEP experiments at CERN one of the main problem was the determination of the number of the flavor neutrinos from the measurement of the invisible width of $Z^{0}$-boson. This problem was successfully solved. One of the main problem of modern neutrino oscillation experiments is the determination of the number of light  neutrinos via the search for transition into sterile states. Hope that this problem will be solved in the nearest years.

The Standard Model teaches us that the simplest possibilities are most likely to be correct. Two-component, left-handed Weyl neutrinos and absence of the right-handed neutrino fields in the Standard Model is the simplest, most elegant and most economical possibility. Majorana  mass term generated by the $L$-violating, dimension five effective Lagrangian  is the simplest, most economical mass term. Future experiments will show whether  this possibility is realized in nature.

We will finish with a relevant citation: ``Simplicity is a guide to the theory choice" A. Einstein.

\section{Acknowledgement}
I would like to thank TRIUMF Theoretical Department for hospitality.
This work was partly supported  by RFBR Grant No. 18-02-00733 A.

\end{document}